\documentclass{article}
\title{A Proposed Casimir-Like Effect Between Contaminants in Ideal Bose-Einstein Condensates}
\author{Alexander Oshmyansky}
\date{January 2007}
\begin{document}
\maketitle

\begin{abstract}
It is hypothesized that, within Bose-Einstein condensates,
contaminants will form a potential that effects the energy state
of a condensate. While assuming a system governed by the
Gross-Pitaevskii equation, contaminants are modelled as boundary
conditions for the wave function of the condensate. It is then
found that the energy of the system depends directly upon the
distance between contaminants. Energy is minimized as two
particles either come together or move apart depending on the
nature of the condensate. This is due to the presence of induced
standing waves in the condensate between two contaminants, similar
to the attractive effect caused by standing electromagnetic waves
in a vacuum, the Casimir effect. Quantum calculations are also
done to determine the expected strength of the ``contaminant in
condensate" effect.
\end{abstract}

\section{Introduction}
\indent Much attention has previously been given to the behavior
of atoms on a macroscopic level within Bose-Einstein condensates
\cite{Pethick02}, especially since their first observation
\cite{Anderson95}. However, relatively little work has been done
on the behavior of contaminants within Bose-Einstein condensates,
such as $\mathrm{He}^{3}$ within a $\mathrm{He}^{4}$ condensate
with the exception of how to eliminate them, for example as
mentioned in \cite{Lewandowski03} and \cite{McGuirk04}. Here the
existence of a ``Casimir-like" effect is proposed within a
Bose-Einstein condensate where the wave function of the condensate
functions as the background electromagnetic field of the classical
Casimir effect \cite{Bordag06}. The proposed effect will be
referred to as the ``contaminant in condensate" (CIC) effect.

\section{Predictions}
The idea of a Casimir-like force between contaminants in
Bose-Einstein condensate is conceptually simple. For the sake of
similar simplicity in our first calculation, the following
assumptions were made.

\begin{flushleft}
\begin{enumerate}

\item Any given contaminant particle can be taken to occupy a
cubic volume element or a point.

\item Tunnelling does not occur through a given contaminant, i.e.
the potential in a contaminant is infinite and can be represented
by an infinite square well.

\item The Bose-Einstein Condensate is ideal.

\item There are no other external potentials.

\item In the absence of any potential, the energy of the wave
function of a Bose-Einstein condensate goes to 0. In general, the
wave function of a Bose-Einstein condensate will always be in the
ground state.

\end{enumerate}
\end{flushleft}
The Gross-Pitaevskii equation representing the state of the
condensate between two contaminants should thus take the form of a
one-dimensional Schr\"{o}dinger wave equation:

\begin{equation}
-\frac{\hbar^{2}}{2m}\nabla^{2}\psi(\textbf{r}) +
V(\textbf{r})\psi(\textbf{r})=\mu\psi(\textbf{r})
\end{equation}

\noindent where $\psi(\textbf{r})$ is the wave function of the
condensed state, \emph{m} is the mass of a given component boson,
and $\mu$ is the chemical potential as usual. $V(\textbf{r})$ is
given in this case by an infinite square well located at the
position of each of the contaminants. This gives a potential for
the condensate between two contaminants of:

\begin{equation}
\mu=\frac{\hbar^{2}\pi^{2}}{8ma^{2}}
\end{equation}

\noindent Note the reason a one-dimensional Schr\"{o}dinger wave
equation is used even in the case of cube-shaped contaminants is
that in directions parallel to the faces of contaminants which are
facing one another, the energy of the wave function of the
condensate is equal to 0, as per assumption 5 above.

\indent The energy of the condensate superstate is inversely
proportional to the square of the distance between the two
contaminants. Taking the negative of the gradient of (2) with
respect to distance $a$ results in a force:

\begin{equation}
F=\frac{\hbar^{2}\pi^{2}}{4ma^{3}}
\end{equation}

\noindent between two contaminants.

To be slightly more sophisticated, relativistic effects can be
taken into account by using the Klein-Gordon equation
\cite{Kaku93}:

\begin{equation}
(\frac{1}{c^2}\frac{\partial^2}{\partial t^2} - \nabla^2
+\frac{m^2c^2}{\hbar^2})\psi=0
\end{equation}

\noindent rather than the Schr\"{o}dinger equation. This has both
positive and negative energy solutions given by:

\begin{equation}
E_n = \pm\sqrt{\frac{m^2c^4}{\hbar^2}+\frac{c^2\pi^2n^2}{a^2}}
\end{equation}

\noindent where $n=1,2,3\ldots$. This begs the question of which
energy level a condensate superstate will naturally settle to. To
resolve this issue, it is assumed that a superstate will obtain
the lowest energy value not otherwise occupied. By assuming some
analogy of the dirac ``electron sea" to exist for this model, it
is speculated that this occurs at either the positive or negative
case of $n=1$.

\subsection{Gross-Pitaevskii Approach}

Let us now drop assumptions 3 and 4 from above and include the
entire Gross-Pitaevskii (GP) equation \cite{Lifshitz78}:

\begin{equation}
-\frac{\hbar^{2}}{2m}\nabla^{2}\psi(\textbf{r},t) +
V(\textbf{r},t)\psi(\textbf{r},t)+U_{0}|\psi(\textbf{r},t)|^{2}\psi(\textbf{r},t)=\mu\psi(\textbf{r},t)
\end{equation}

\noindent  to allow calculations in a more physically realistic
model where $U_{0}=\frac{4\pi\hbar^2\alpha}{m}$ gives the
interaction, attractive or repulsive, between two atoms with
scattering radius $\alpha$. In this case $V(\textbf{r})$ is given
by:

\begin{equation}
V(x,y,z)=\frac{1}{2}m(\omega^{2}x^{2}+\omega^{2}y^{2}+\omega^{2}z^{2})+\sum_{j}B(x,y,z,x_1,y_1,z_1),
\end{equation}

\noindent where the function $B$ is given for each contaminant $j$
by:

\begin{equation}
B(x,y,z,x_1,y_2,z_2)=\cases{\infty &if $x=x_1$, $y=y_1$,
$z=z_1$\cr
              0 &if otherwise\cr}
\end{equation}

\noindent and $x_1, y_1,$ and $z_1$ (or alternatively
$\mathbf{r_1}$) is the position of each contaminant $j$.

\indent There is no analytic solution to the Gross-Pitaevskii
equation. For an approximately solution, the technique of Edwards
and Burnett \cite{Edwards95} will be used. It will be assumed that
the kinetic energy of the condensate is significantly less than
the external potential and contribution of interatomic forces. The
GP equation can thus be changed to:

\begin{equation}
(\frac{1}{2} m \omega^2 r^2 + \sum_j B(\mathbf{r,
r_j}))\psi(\textbf{r}) + N U_0
|\psi(\textbf{r})|^2\psi(\textbf{r})=\mu \psi(\textbf{r}),
\end{equation}

\noindent the solution of which, using the factor $\sum_j
B(\mathbf{r,r_j})$ as boundary, conditions is:

\begin{equation}
|\psi(\mathbf{r})|^2=\frac{\mu-(1/2)m\omega^2r^2}{NU_0}.
\end{equation}

This has the advantage of going to zero as $r$ goes to a critical
value $r_c$, thus fitting well to the wavefunction for a
condensate between two infinite point potentials. The potential
associated with this wavefunction is:

\begin{equation}
\mu=\frac{m\omega^2r^2_c}{2}.
\end{equation}

By replacing $r_c$ with $a/2$, we get a force associated with the
presence of a condensate superstate between two contaminants of:

\begin{equation}
F(a)=-\frac{m\omega^2a}{4}
\end{equation}

\noindent Note the rest of the condensate, disregarding the effect
of atom density, is unaffected by the distance between the two
contaminants and thus has no effect on the CIC effect.

\subsection{Quantum Approach}

Finally, by taking a quantum approach, as used in the Bogoliubov
approach \cite{Bogoliubov47}, rather than a classical approach, we
can find a way to eliminate assumption 5 and take into account an
entire excitation spectrum of an ideal condensate. Given the
nature of a Bose-Einstein condensate, assumption 5 is likely to be
physically valid. However, in the interest of thoroughness, we
will show how it can be disregarded here. The usual derivation of
the Casimir effect shall be used and will be recounted here
\cite{Bordag06}. We shall assume that a Bose-Einstein condensate
is governed by the massive scalar field equation:

\begin{equation}
\frac{1}{c^2}\frac{\partial^2\varphi(t,x)}{\partial
t^2}-\frac{\partial^2\varphi(t,x)}{\partial x^2} +
\frac{m^2c^2}{\hbar^2}\varphi(t,x)=0
\end{equation}

\noindent Solutions of this equation are given by:

\begin{equation}
\varphi^{(\pm)}_n(t,x) = \sqrt{\frac{c}{a \omega_n}} e^{\pm i
\omega_n t} \sin k_nx
\end{equation}

\begin{equation}
 \omega_n = \sqrt{\frac{m^2c^4}{\hbar^2}+c^2k_n^2}
\end{equation}

\begin{equation}
 k_n=\frac{\pi n}{a}, n = 1,2,...
\end{equation}

We now quantize this field by using the expansion:

\begin{equation}
\varphi(t,x)=\sum_n[\varphi^{(-)}_n(t,x)a_n +
\varphi^{(+)}_n(t,x)a_n^+]
\end{equation}

\noindent with commutation relations for the annihilation and
creation operators

\begin{equation}
[a_n,a_{n'}^+]=\delta_{n,n'}, [a_n,a_{n'}]=[a_n^+,a_{n'}]=0
\end{equation}

\noindent and vacuum state defined by:

\begin{equation}
a_n|0\rangle = 0.
\end{equation}

\noindent The desired quantity is the energy density:

\begin{equation}
T_{00}(x)= \frac{\hbar c}{2}(c^{-2}[\partial_t\varphi(x)]^2 +
[\partial_x\varphi(x)]^2)
\end{equation}

\noindent which from equations (12)-(18) is:

\begin{equation}
\langle 0 | T_{00}(x) | 0 \rangle = \frac{\hbar}{2a}
\sum_{n=1}^{\infty}\omega_n - \frac{m^2c^4}{2a\hbar}
\sum_{n=1}^{\infty}\frac{\cos 2k_nx}{\omega_n}.
\end{equation}

\noindent Total energy across an interval length $a$ is:

\begin{equation}
E(a)\int^a_0 \langle 0 | T_{00}(x) | 0 \rangle dx =
\frac{\hbar}{2}\sum_{n=1}^{\infty}\omega_n.
\end{equation}

\noindent This value is evaluated by introducing a damping
function, with the result:

\begin{equation}
E(a) = -\frac{mc^2}{4}-\frac{\hbar c}{4\pi a}\int_{2\mu}^\infty
\frac{\sqrt{y^2-4}\mu^2}{e^y-1}dy
\end{equation}

\noindent where $\mu \equiv mca/\hbar$. For $\mu \gg 1$ as is the
case in our calculations:

\begin{equation}
E(a)\approx -\frac{mc^2}{2} - \frac{\sqrt{\mu}\hbar
c}{4\sqrt{\pi}a}e^{-2\mu}.
\end{equation}

It should also be noted that the quantum approach gives another
rather convenient method for calculating the force of attraction
between two contaminants due to a single condensate superstate. To
do so we associate with the superstate a scalar propagator for a
spin 0 particle:

\begin{equation}
\frac{1}{k^2-m^2+i\varepsilon}.
\end{equation}

If we then let the superstate ``propagate" from one distance to
another rather than from one position to another, one can use
Feynman's path integral approach to calculate the energy and force
associated with the creation and propagation of a superstate. This
turns out to be, using the usual methods \cite{Zee03}:

\begin{equation}
E(a)= -\frac{1}{4\pi a}e^{-ma}.
\end{equation}

\section{Discussion}
\indent It is interesting to note the similarities of the above
effect to the Casimir effect in a vacuum. In the Casimir effect,
standing waves due to an electromagnetic field in a vacuum result
in an attractive force between any two objects in what would
otherwise be a vacuum. However, the Casimir effect requires
summation over all possible excitation modes of the standing
waves. As we assume only one excitation mode for a Bose-Einstein
condensate, we remove that step in the computation.

\indent The pressing issue that occurs is whether the above
hypothesis actually could be observed to occur. It is clear that
this would be difficult, as contaminants obviously tend to prevent
a Bose-Einstein condensate from forming and destabilize the
condensate around them. In addition, the strength of the proposed
interaction is so small that it would be very difficult to
directly observe for systems as small as a few dozen atoms large.

Rather than direct observation, an alternative approach is to note
that the potential energy that forms a condensate supersate must
come from the kinetic energy of its components. The introduction
of contaminants into a Bose-Einstein condensate which do not
couple to it and that are cooled to the level of the surrounding
condensate should thus paradoxically lower the temperature of the
surrounding condensate. The reduction in temperature should
proportional to the energy of the induced condensate superstates.

\section{Acknowledgements}
I would like to thank P.K. Maini as well as the rest of the
students and faculty at his lab group and at the Maths Institute
at Oxford. I would also like to thank Keith Burnett for his input
and early review of this paper. Funding for this project was
provided by the Marshall Aid Commemoration Commission.

\end{document}